\documentclass[fleqn,usenatbib]{mnras}

\usepackage{newtxtext}

\usepackage[T1]{fontenc}

\DeclareRobustCommand{\VAN}[3]{#2}
\let\VANthebibliography\thebibliography
\def\thebibliography{\DeclareRobustCommand{\VAN}[3]{##3}\VANthebibliography}

\usepackage{graphicx}	
\usepackage{amsmath}	
\usepackage{amssymb}	
\usepackage{float}
\usepackage{soul}

\newcommand{\coleman}[1]{\textcolor{cyan}{#1}}

\title[Cosmology from Large Populations of Strong Lenses]{Cosmology from Large Populations of Galaxy-Galaxy Strong Gravitational Lenses}

\author[Tian Li et al.]{
Tian Li\thanks{tian.li@port.ac.uk},
Thomas E. Collett,
Coleman M. Krawczyk
and Wolfgang Enzi
\\
Institute of Cosmology and Gravitation, University of Portsmouth, Burnaby Rd, Portsmouth PO1 3FX, UK\\
}

\date{Accepted XXX. Received YYY; in original form ZZZ}

\pubyear{2023}

\begin{document}
\label{firstpage}
\pagerange{\pageref{firstpage}--\pageref{lastpage}}
\maketitle

\begin{abstract}
We present a forecast analysis on the feasibility of measuring the cosmological parameters with a large number of galaxy-galaxy scale strong gravitational lensing systems. Future wide area surveys are expected to discover and measure the properties of more than 10 000 strong lensing systems. We develop a hierarchical model that can simultaneously constrain the lens population and cosmological parameters by combining Einstein radius measurements with stellar dynamical mass estimates for every lens. Marginalizing over the lens density profiles and stellar orbital anisotropies, we find that $w$ can be constrained to a precision of $0.11$ with 10 000 galaxy-galaxy lens systems, which would be better than any existing single-probe constraint. We test our method on 161 existing lenses, finding $w=-0.96\pm0.46$. We also show how to mitigate against the potential systematic of redshift evolution in the mean lens density profile of the population.

\end{abstract}




\section{Introduction}

The acceleration of the universe has been discovered through Type-Ia supernovae \citep{1999ApJ...517..565P,1998AJ....116.1009R}, and has been measured by several observational probes, including Cosmic Microwave Background (CMB) Anisotropies, Baryon Acoustic Oscillations, Weak Gravitational Lensing, Galaxy Clustering, and Redshift Space Distortion \citep{2014arXiv1401.0046M}. These observations concluded that under flat $\Lambda$CDM model, the dark energy density makes up 70$\%$ of the universe today, and has an equation of state of $w \approx -1$. However, the exact nature of dark energy and dark matter still remains unknown. The 5 $\sigma$ discrepancy of Hubble Constant between Planck's CMB measurement \citep{2014A&A...571A..16P,2016A&A...594A..13P,2020A&A...641A...6P} and local measurements of supernovae \citep{2011ApJ...730..119R, 2012ApJ...758...24F, 2016ApJ...826...56R} also suggest potential new physics beyond the $\Lambda$CDM model.

\begin{figure}
    \centering
    \includegraphics[width=0.5\textwidth]{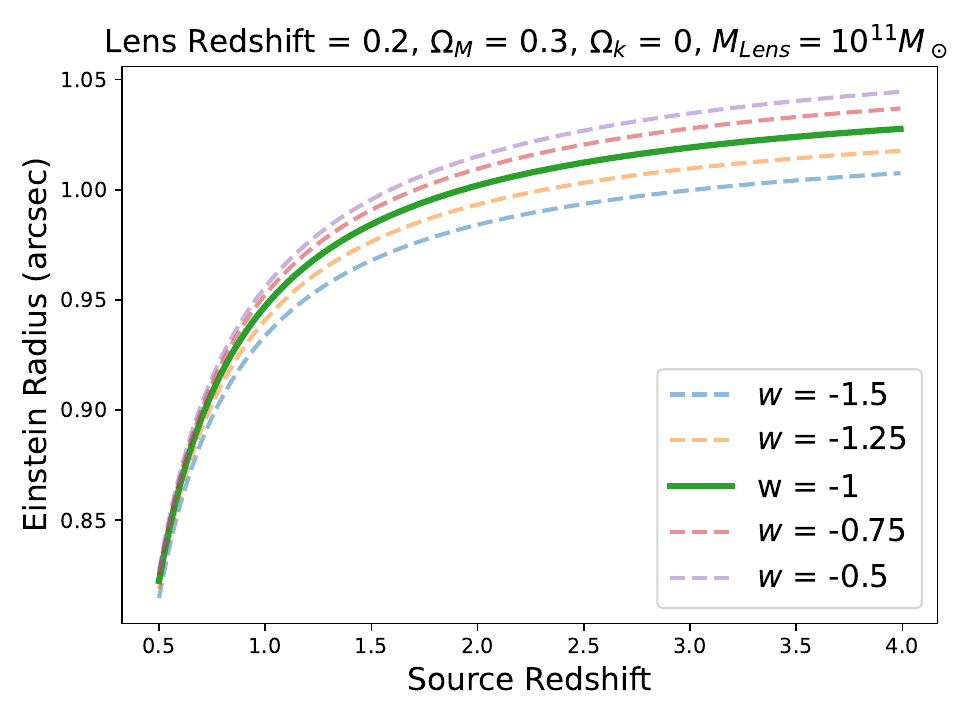}
    \caption{The Einstein radius as a function of the source redshift, for different values of the equation of state of dark energy. All other parameters are fixed. We assume the redshift of lens galaxy is 0.2, and has an Einstein mass of $10^{11}\mathrm{M}\odot$.}
    \label{fig:rein-w}
\end{figure}

In addition to the above methods, galaxy scale strong gravitational lensing provides an independent probe to constrain the cosmological parameters. Strong gravitational lensing occurs when two galaxies align perfectly to our line of sight, such that the light from the background source galaxy will be distorted and magnified by the foreground lens galaxy, resulting in multiple sources or an Einstein ring in the observed image \citep{1936Sci....84..506E, PhysRev.51.290, PhysRev.51.679}. The radius of the Einstein ring (Eintein radius), relative positions, flux ratios, and time delays between multiple images depend both on the gravitational potential of the lens galaxy and angular diameter distances between the observer, lens galaxy, and source galaxy. It is the dependence on the angular diameter distances that makes strong lensing sensitive to cosmological parameters. Figure \ref{fig:rein-w} illustrates the sensitivity of the Einstein radius to the equation of state of dark energy. The most well-studied method of constraining cosmology with strong lensing is time-delay cosmography, which uses the temporal variation of a gravitationally lensed quasar or supernova to constrain the Hubble constant \citep{Refsdal:1964nw, 2022arXiv221010833B,2016A&ARv..24...11T,2022A&ARv..30....8T}. Alternatively, the equation of state of dark energy, $w$, can be measured from systems with sources at multiple redshifts \citep{gavazzi2008}: \citet{2014MNRAS.443..969C} used a single double source plane lens to infer $w=-0.99_{-0.22}^{+0.19}$. Aside from single lens analyses, statistical analyses of the ensemble of lens systems can also place bounds on cosmological parameters. \cite{SDSS:2007jtw}, \cite{PhysRevLett.89.151301}, and \cite{Chae:2006kw} studied lenses from Cosmic Lens All-sky Survey  \citep[CLASS,][]{2003MNRAS.341....1M} and The Sloan Digital Sky Survey Quasar Lens Search  \citep[SQLS,][]{2012AJ....143..119I}. They measured $w$ through comparing the empirical distribution of image separations in observed samples of lenses with theoretical models. Since the number of lens systems is low, the constraint on $w$ using this method is weak (e.g., $w=-1.1 \pm 0.6_{-0.5}^{+0.3}$, \cite{SDSS:2007jtw}).

The current cosmology analyses with strong lensing systems are limited by several factors.
The primary issue is that the sample of known galaxy-scale lenses is only a few hundred systems, discovered in several surveys with heterogeneous selection functions. Once a sample of time-delay or compound lenses has been selected only a handful of suitable systems have adequate data for precision cosmography, e.g the latest TDCOSMO sample has only 6 time-delay lenses \citep{2020A&A...643A.165B}. Accurate and efficient lens modelling is also a challenge. The mass distribution in the lens must be inferred to convert lensing observables into cosmological constraints. This challenge is compounded by the mass-sheet degeneracy, where different mass models can produce identical strong lensing observables but imply different cosmological parameters \citep{Schneider&Sluse2013}. 

Future optical imaging surveys, including Euclid and LSST, are predicted to discover more than $10^6$ galaxy-galaxy strong lensing systems \citep{2015ApJ...811...20C}. Euclid will provide high-resolution images, such that lens modelling could plausibly be performed for every lens, without the need for additional imaging data. However, almost all strong lensing science requires accurate lens and source redshifts, and the 4MOST Strong Lensing Spectroscopic Legacy Survey (4SLSLS) is expected to obtain the spectrum of $\approx$ 10 000 strong lensing systems \citep{2023Msngr.190...49C}.

Combining lensing and stellar dynamics opens up a new statistical method to constrain cosmology. Gravitational lensing determines the mass inside Einstein Radius, and stellar kinematics determines the gravitational potential within which the stars are moving \citep{Koopmans:2005ig}. Assuming general relativity, the dynamic mass enclosed with the Einstein radius must equal the gravitational mass measured using the Einstein radius. 

Thus the combination of lensing and dynamical observables are sensitive to the mass profile of the lens, the orbital properties of its stars, and the cosmological distances \citep{Futamase:2000hnr, Grillo:2007iv}.  The challenge inherent to this method is that the mass profile of lenses and the orbital profiles of their stars are not well known. To produce cosmological constraints, either assumptions must be made about the lenses, or the cosmological parameters and lens properties must be inferred simultaneously. \cite{2010MNRAS.406.1055B} applied the lensing plus dynamics method on 20 lens systems. By assuming a SIS mass profile for all lens galaxies, they found $\Omega_M$ = 0.27 $\pm$ 0.28, and $w = -0.63 \pm 0.45$. \cite{Cao_2015} and  \cite{Chen:2018jcf} improved on this by fitting $\sim 100$ lenses with powerlaw density profiles.

In this paper, we investigate how well the combination of Euclid and 4MOST data for 10 000 lenses can constrain the cosmological parameters. We employ a Bayesian hierarchical model and simultaneously fit for the cosmological parameters and the ensemble properties of the lens galaxies, including the density profile slope and the stellar orbital anisotropy. It is hard to perform detailed modelling for 10 000 lens systems, so we assume that we can only use catalog-level data in our analysis.  

The rest of the paper is organized as follows. In Section 2, we present the mass model of the lens galaxy and the equation that relates the galaxy's velocity dispersion to cosmological parameters. We then introduce the properties of the mock data and discuss potential future surveys for data acquisition. In Section 3, we describe the hierarchical model used to simultaneously fit the lens galaxy properties and cosmology. In Section 4, we present the results obtained under different cosmological models and data measurement accuracies. The final section summarizes the main conclusions. In this paper, the fiducial cosmology for the mock data set is as follows: $\Omega_\mathrm{M}=0.3$, $\Omega_\Lambda=0.7$, $\Omega_k=0$, and $w=-1$.

\section{Theory}

The mass enclosed within the Einstein ring can be related to Einstein radius through:
\begin{equation}
\theta_{\mathrm{E}}=\sqrt{\frac{4 G M\left(\theta_{\mathrm{E}}\right)}{c^2} \frac{D_{\mathrm{ls}}}{D_{\mathrm{l}} D_{\mathrm{s}}}}
\end{equation}
 where $D_l$ is the angular diameter distance of the lens galaxy, $D_{ls}$ is the angular diameter distance between lens and source, and $D_{s}$ is that between observer and source. $\theta_E$ is the angular Einstein radius. M\coleman{(}$\theta_\mathrm{E}$) is the galaxy mass enclosed within the Einstein radius, and the angular diameter distance is :
\begin{eqnarray}
D_{\mathrm{ij}}=\frac{c / H_0}{\left(1+z_{\mathrm{j}}\right)}\left(\frac{\operatorname{sinn}\left(\sqrt{\left|\Omega_k\right|} \int_{z_{\mathrm{i}}}^{z_{\mathrm{i}}} \frac{\mathrm{d} z}{E(z)}\right)}{\sqrt{\left|\Omega_k\right|}}\right)
\end{eqnarray}
where sinn(x) = sin(x), x, or sinh(x) for open ($\Omega_k < 0$), flat ($\Omega_k = 0$), or closed ($\Omega_k > 0$) universes respectively, and E(z) is the normalised Hubble parameter:
\begin{eqnarray}
E(z)=\sqrt{\Omega_{\mathrm{M}}(1+z)^3+\Omega_k(1+z)^2+\Omega_\Lambda(1+z)^{3(1+w)}}
\end{eqnarray}

When combining lensing and dynamics, the cosmological model is not directly probed by the measurement of a single distance, but instead through a ratio of distances $\frac{D_s}{D_{l s}}$. However, to make this measurement we need a model that can connect the dynamical data with the lensing mass. Since lenses are typically elliptical galaxies (ETGs) with E/S0 morphologies \citep{2010MNRAS.405.2579O}, we use a power profile for both total mass density and stellar luminosity profile \citep{Koopmans:2005ig}:
\begin{eqnarray}
\begin{array}{l}
\rho(r)=\rho_0\left(\frac{r}{r_0}\right)^{-\gamma} \\
\nu(r)=\nu_0\left(\frac{r}{r_0}\right)^{-\delta} \\
\beta(r)=1-\frac{\sigma_\theta^2}{\sigma_r^2}
\end{array}.
\end{eqnarray}
where $\rho(r)$ is the mass density (include dark matter) distribution function. $\nu(r)$ is the luminosity density of stars. $\beta(r)$ is the anisotropy of the stellar velocity dispersion (stellar orbital anisotropy), where $\sigma_\theta$ and $\sigma_r$ are the tangential and radial velocity dispersion, respectively. $\beta$ ranges from +1 to $-\infty$, where $\beta=0$ corresponds to the "isotropic" case, $\beta=1$ corresponds to a galaxy with pure circular stellar movement, $\beta= -\infty$ means that stars in the galaxy only have radial movement. In the scenario where $\gamma=\delta=2$ and $\beta=0$, the mass model reduces to the Singular Isothermal Sphere (SIS) model, which is a commonly used approximation for the mass profiles of elliptical galaxies (e.g.  \cite{2010ApJ...724..511A}).

After solving the spherical Jeans equation and substituting the dynamical mass into  equation (2) we get \citep{Koopmans:2005ig}: 
\begin{equation}
\label{eqa:main}
\sigma_{\|}^2(R_A)=\frac{c^2}{2 \sqrt{\pi}} \frac{D_s}{D_{l s}} \theta_E \times f\left(\gamma, \delta, \beta\right) \left(\frac{\theta_A}{\theta_E}\right)^{2-\gamma},
\end{equation}
where
\begin{equation}
\label{eqa:f}
\begin{array}{r}
    f\left(\gamma, \delta, \beta\right) = \frac{3-\delta}{(\xi-2 \beta)(3-\xi)} 
    \left[\frac{\Gamma[(\xi-1) / 2]}{\Gamma(\xi / 2)}-\beta \frac{\Gamma[(\xi+1) / 2]}{\Gamma[(\xi+2) / 2]}\right] \\
    \times \frac{\Gamma(\gamma / 2) \Gamma(\delta / 2)}{\Gamma[(\gamma-1) / 2] \Gamma[(\delta-1) / 2]}
\end{array}
\end{equation}
$\sigma_{\|}^2(R_A)$ is the luminosity averaged line-of-sight velocity dispersion (LOSVD) measured in a circular fibre of radius $\theta_A$ \footnote{For 4MOST the fibre diameter is 1.45 arcseconds \citep{2012SPIE.8446E..0TD}.}. $\Gamma$s are Gamma functions, and $\xi=\gamma+\delta-2$.  

\begin{figure}
    \centering
    \includegraphics[width=0.55\textwidth]{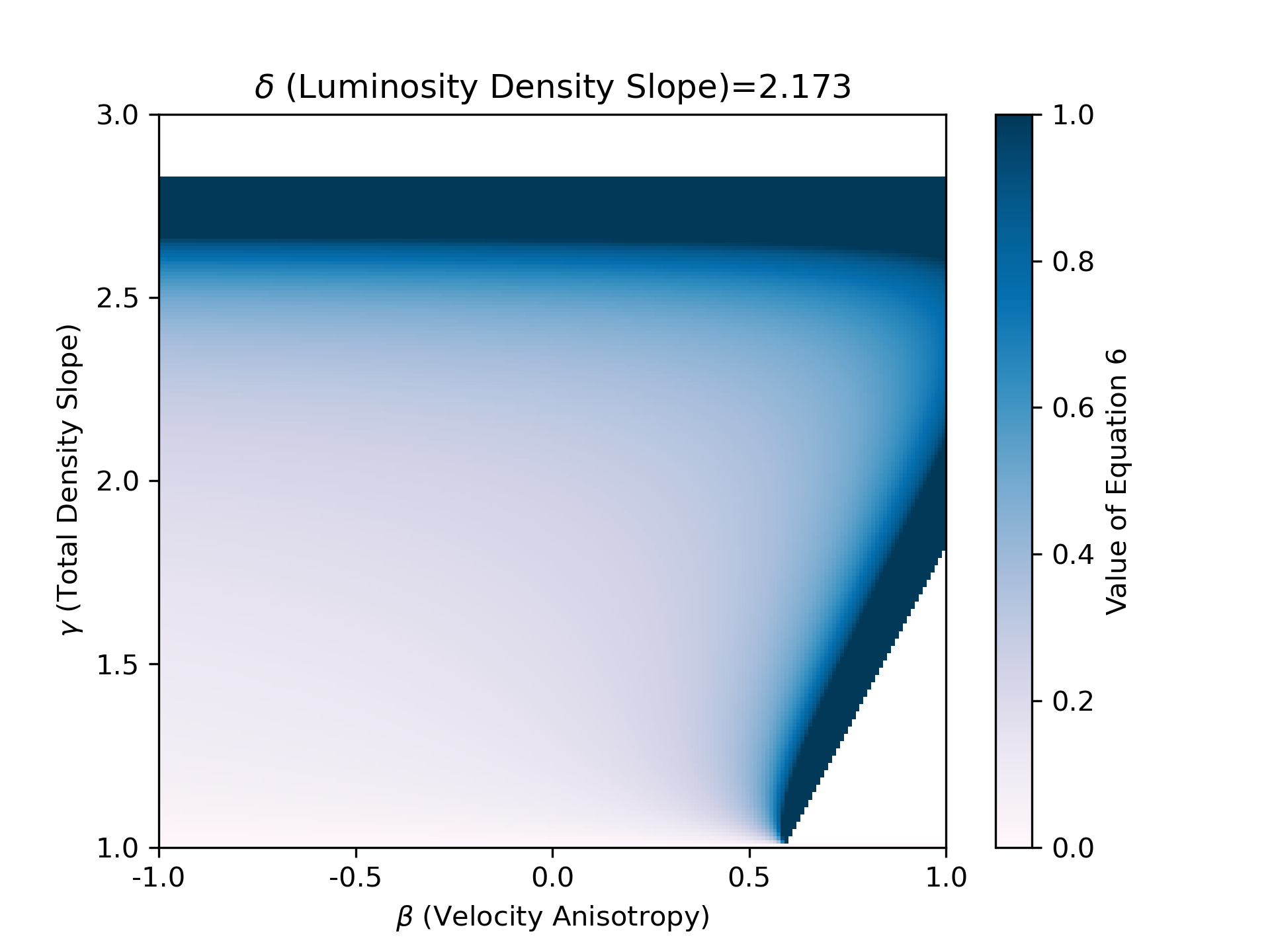}
    \caption{The value of the function that links Einstein radius and velocity disperion as a function of the orbital anisotropy and lens density profile slope. It is given in equation \ref{eqa:f} as a function of $\gamma$ and $\beta$. We have fixed $\delta=2.173$. The white regions are unphysical as the luminosity-weighted mass diverges here.}
    \label{fig:ffunc}
\end{figure}

Solving Equation \ref{eqa:main}, we find that for a fixed galaxy mass and surface brightness profile, a steeper density profile (larger $\gamma$) and a higher velocity anisotropy (larger $\beta$) lead to a higher stellar velocity dispersion. Figure \ref{fig:ffunc} shows the value of Equation \ref{eqa:f} as a function of $\gamma$ and $\beta$, assuming $\delta = 2.173$, which is the typical value of SLACS lenses. 

\begin{figure*}
    \centering
    \includegraphics[width=0.7\textwidth]{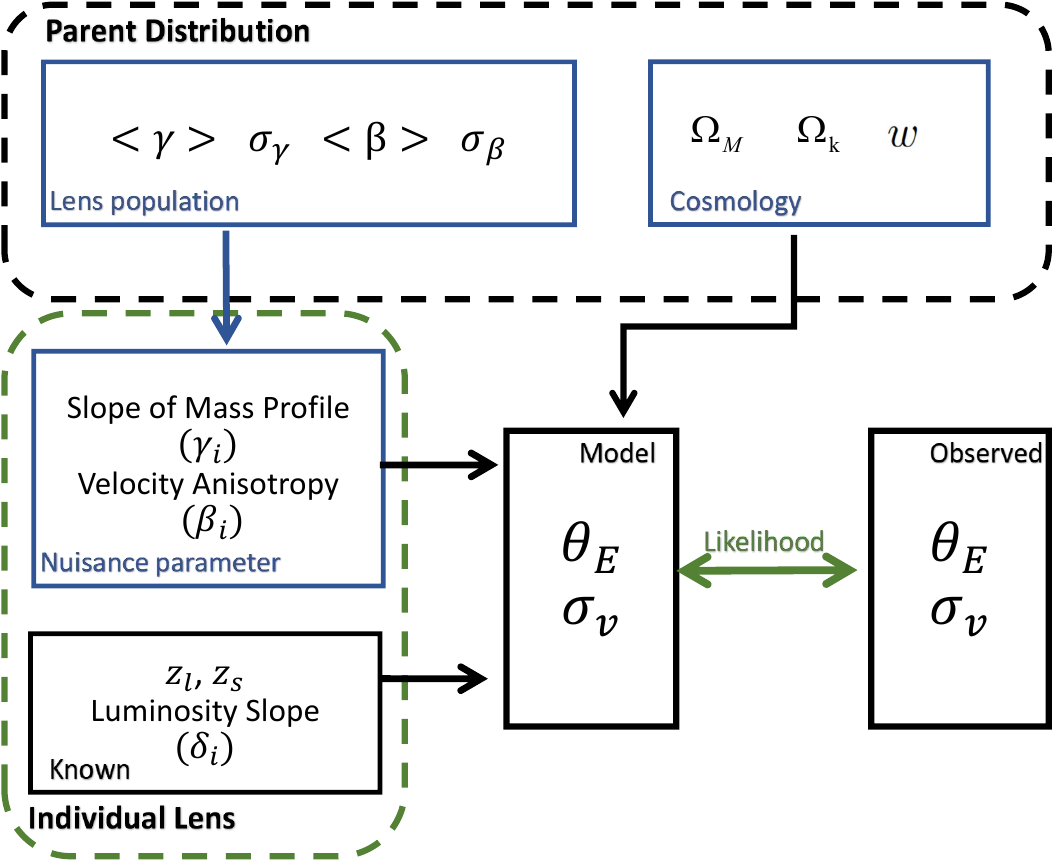}
    \caption{Schematic of our hierarchical model. Solid blue boxes are parameters of our model, and solid black boxes are observables or predictions of our model. 
    The black dashed box includes the hyperparameters of the model, the parent lens population parameters and the cosmological parameters. The green dashed box contains the properties of individual lens systems, including the nuisance parameters $\gamma_i$ and $\beta_i$ of individual lens galaxies, which are drawn from the population parent distribution. Modeled velocity dispersions and Einstein radii can be computed through combining the parameters of the individual lenses and cosmology. Comparing these modelled values to the observed values, a likelihood function can be built up, which we sample with a Hamiltonian Monte Carlo analysis.}
    \label{fig:model}
\end{figure*}

\subsection{Mock data}
To build our mock sample of lenses, we use the simulated strong lensing population forecast to be observed by Euclid from \cite{2015ApJ...811...20C}. We remove all systems with a source redshift greater than 1.5 since the [OII] doublet is redshifted out of the 4MOST wavelength range at this redshift. The sample is built assuming lenses are uniformly distributed in co-moving volume, follow the observed velocity dispersion function of SDSS \citep{Choi:2006qg}, and have an SIS density profile. The source properties use the LSST sky simulations, with redshifts and number counts matched to observations \citep{Connolly2010}. These assumptions produce a realistic lens population, but lack the complexities of a non-SIS density profile, or of stellar anisotropy. Therefore, we take only the lens galaxy redshift, source galaxy redshift, and the Einstein radius in the data set. We assign $\gamma$, $\beta$, and $\delta$ values to each lens galaxy to produce new velocity dispersions. 

Observations shows that $\gamma$ has a distribution of $2.078 \pm 0.16$ \citep{2010ApJ...724..511A}, $\delta$ has a distribution of $2.173 \pm 0.085$ \citep{Chen:2018jcf}, and $\beta=0.18 \pm 0.13$ \citep{2010ApJ...708..750S, 2006PhRvD..74f1501B}. We generate mock galaxies using these distributions. The true values of velocity dispersion are generated through equation (6). Then, we add random noise to these values to simulate measurement error.
The measurement error of Einstein radius is set as 0.01 arcsec, and the error on velocity dispersion of 4MOST is 10km/s. We neglect errors on $\delta$ and the redshifts since they can be measured with high accuracy. In our fiducial model, we assume that $\beta$ and $\gamma$ for each lens are unknown. The full setup of mock data is shown in Table \ref{tab:mock}.

\begin{table}
  \centering
  \begin{tabular}{|c|l|c|c|}
    \hline
    \textbf{Property name} & \textbf{Distribution } & \textbf{Measurement error} \\
    \hline
    $\gamma$ & 2 $\pm$ 0.16  & 0.02* \\
    $\delta$ & 2.173 $\pm$ 0.085 & None  \\
    $\beta$  & 0.18 $\pm$ 0.13  & $-$\\
    $\sigma_v$ & Equation \ref{eqa:main}  & 10 km/s\\
    $\theta_\mathrm{E}$ & \citet{2015ApJ...811...20C}  & 0.01''\\
    $z_l$, $z_s$  & \citet{2015ApJ...811...20C}  & None\\
    \hline\\
    \multicolumn{3}{l}{*$\gamma$ is only treated as an observable in Section \ref{sec:evolving}}\\
  \end{tabular}
  \caption{The parameters of the mock strong lensing systems used in this work. Einstein radius and redshifts are taken from \citet{2015ApJ...811...20C}, with source redshifts below $z_s < 1.5$. The velocity dispersions are computed from the other parameters using equation \ref{eqa:main}. Among the above properties, $\delta$, redshift, Einstein radius, and Velocity dispersion are treated as observables. The measurement error of $\delta$ and redshift are negligible. $\gamma$ can also be measured through detailed lens modelling, but we treat it as a free parameter except in the end of Section \ref{sec:evolving}.}
  \label{tab:mock}
\end{table}

\section{Hierarchical Model}

Since we do not know the mass profile or orbital anisotropy of individual lenses apriori, and they are not easily measured without detailed lens modelling and integral field unit kinematics, we instead require a hierarchical model to connect the measured Einstein radius and the velocity dispersion of every lens with the underlying cosmological parameters of the Universe. In fact, even the population properties for lens density profiles and stellar anisotropies are not well known once the cosmology is allowed to be a free parameter. For our population model, we assume that density profile slopes and stellar anisotropies follow a Gaussian distribution. Our hierarchical model must therefore fit for the ensemble mean density profile slope, $\langle\gamma\rangle$, and scatter, $\sigma_\gamma$, the ensemble mean anisotropy $\langle\beta\rangle$, and scatter, $\sigma_\beta$, the individual slopes, $\gamma_i$, and anisotropies, $\beta_i$, of each lens, and the underlying cosmological model parameters. Figure \ref{fig:model} illustrate the structure of our hierarchical model, with the observed Einstein radii and velocity dispersions linked to the model parameters through equation \ref{eqa:main}.

\subsection{Sampling the parameters of our hierarchical model}

Our hierarchical model has a large number of free parameters ( $\sim$ 3 per observed lens $\times$ 10 000 lenses), traditional MCMC methods are not sufficient for exploring such a posterior distribution.  To that end, we make use of the computational frameworks initially developed for large machine learning models, specifically the NumPyro \citep{phan2019composable, bingham2019pyro} probabilistic programming language (PPL).

NumPyro is an extension of the Pyro \citep{bingham2019pyro} framework that uses JAX\footnote{\url{http://github.com/google/jax}}
for automatic differentiation and adds various Hamiltonian Monte-Carlo (HMC) \citep{1987PhLB..195..216D, brooks2011handbook} sampling methods.  Automatic differentiation is a technique that efficiently computes the partial derivative of a function without the need to write any new code \citep{Baydin2018}, allowing for the computation of gradients of the likelihood function to be evaluated for all free parameters in the model.

We estimate the posterior distribution through NumPyro's No-U-Turn-Sampler (NUTS).  This is an HMC sampler that uses gradient information to find new draw proposals from the likelihood.  Its main advantages over traditional MCMC methods are it produces chains with small auto-correlation values (e.g. each draw is independent), requires relatively few warm-up steps, and can draw samples from very high dimensional distributions without issue.

The priors of the model parameters are as follows:
\begin{equation*}
\begin{aligned}
\text { Cosmology }&\text{: }\left\{\begin{aligned}
\Omega_\mathrm{M} &\sim \mathcal{U}(0,1)\\
\Omega_k &\sim \mathcal{U}(-1,1) \\
w &\sim \mathcal{U}(-2,0)\\
w_a &\sim \mathcal{U}(-3, 1)\\
w_0 &\sim \mathcal{U}(-2,2)
\end{aligned}\right.\\
\text { Parent Lens Distribution }&\text{: }\left\{\begin{aligned}
\text { }\gamma &\sim \mathcal{U}(1.5,2.5)\\
\text { }\sigma_\gamma &\sim \mathcal{HN}(0.5) \\
\text { }\beta &\sim \mathcal{U}(-0.6,1)\\
\text { }\sigma_\beta &\sim \mathcal{HN}(0.5)\\
\end{aligned}\right.\\
\text {Individual Lenses}&\text{: }\left\{\begin{aligned}
\text {   }\gamma_i &\sim \mathcal{U}(\gamma,\sigma_\gamma)\\
\text {   }\beta_i &\sim \mathcal{U}(\beta,\sigma_\beta)\\
\end{aligned}\right.
\end{aligned}
\label{eqa:prior}
\end{equation*}

Where $\mathcal{U}$ stands for uniform distribution, {and} $\mathcal{HN}$ is half normal distribution. Observables like Einstein radius and velocity dispersion have a Gaussian prior centered on their observed values and have a standard deviation that comes from the observational uncertainty.  {For NumPyro's NUTS sampler} we use "$init\_to\_median$" as the initialization strategy with $target\_accept\_prob$ = 0.99. We run 10 chains with 1000 warm-up steps and 1500 sampling steps. Under this set-up, a typical HMC run takes less than 1 hour on an A100 GPU. The typical $\hat{r}$ {for each fit parameter} is smaller than 1.05 with no divergent samples, indicating that all the chains independently converged on the same values.

\section{Results}
\label{sec:results}
\begin{figure*}
    \centering
    \includegraphics[width=\textwidth]{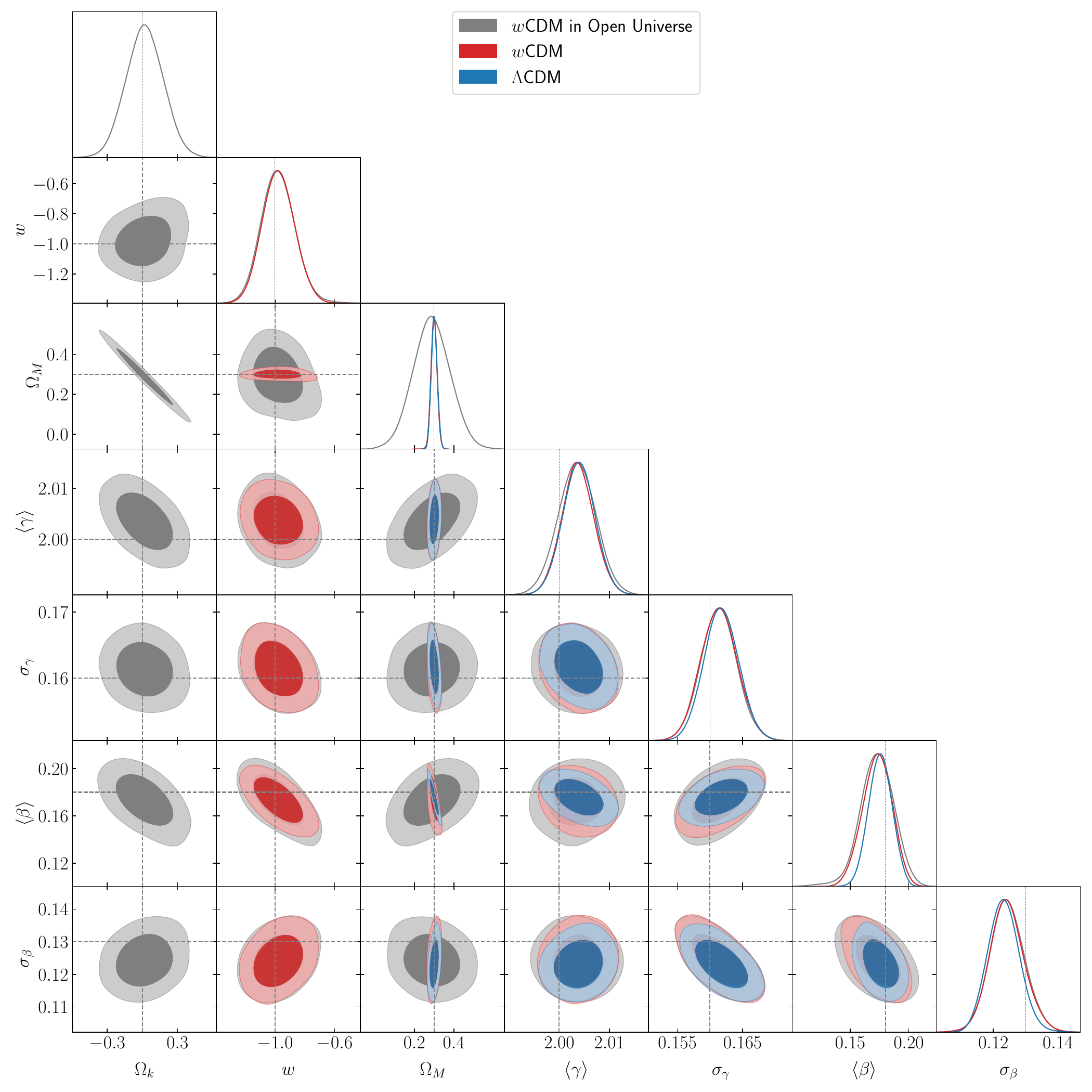}
    \begin{tabular}{c|c|c|c|c|c|c|c|c}
        \hline
        Parameter & $\Omega_M$ & $w$ & $\Omega_k$ & $\gamma$ & $\sigma_{\gamma}$ & $\beta$ & $\sigma_{\beta}$ \\
        \hline
        Fiducial & 0.3 & -1 & 0 & 2 & 0.16 & 0.18 & 0.13 \\
        \hline
        $\Lambda$CDM & 0.301 $\pm$ 0.015 & - & - & 2.004 $\pm$ 0.003 & 0.162 $\pm$ 0.003 &0.176 $\pm$ 0.010 &0.123 $\pm$ 0.005 \\
        $w$CDM & 0.300 $\pm$ 0.015 & -0.98 $\pm$ 0.11 & - & 2.004 $\pm$ 0.003 & 0.161 $\pm$ 0.003 & 0.173 $\pm$ 0.012 & 0.124 $\pm$ 0.005\\
        o$w$CDM & 0.288 $\pm$ 0.094 & -0.98 $\pm$ 0.11 & 0.02 $\pm$ 0.16 & 2.004 $\pm$ 0.004 & 0.161 $\pm$ 0.003 & 0.173 $\pm$ 0.015 & 0.124 $\pm$ 0.005\\
        \hline
    \end{tabular}
    \caption{Posterior distributions and 1D marginalized posterior distributions of the cosmological parameters and the lens population parameters for 10 000 lens systems. Grey assumes an o$w$CDM universe, red is for $w$CDM, and blue is for $\Lambda$CDM universe. $\gamma$ and $\sigma_{\gamma}$ are the population density profile slope mean and standard deviation respectively. $\beta$ and $\sigma_{\beta}$ are the mean and standard deviation values of the population's velocity anisotropy. The contour shows the 68$\%$ and 95$\%$ confidence levels, while the grey dashed line represents the fiducial value which was used to generate our mock data.}
    \label{fig:cdmcorner}
\end{figure*}

We apply our hierarchical approach to model up to 10 000 lens systems and fit three different cosmological models. Aside from the $\Lambda$CDM model, we fit the $w$CDM model, where the equation of state of dark energy is not fixed at -1. We also fit the o$w$CDM model, which allows the universe to have positive or negative curvature. 

\begin{figure}
    \centering
    \includegraphics[width=0.5\textwidth]{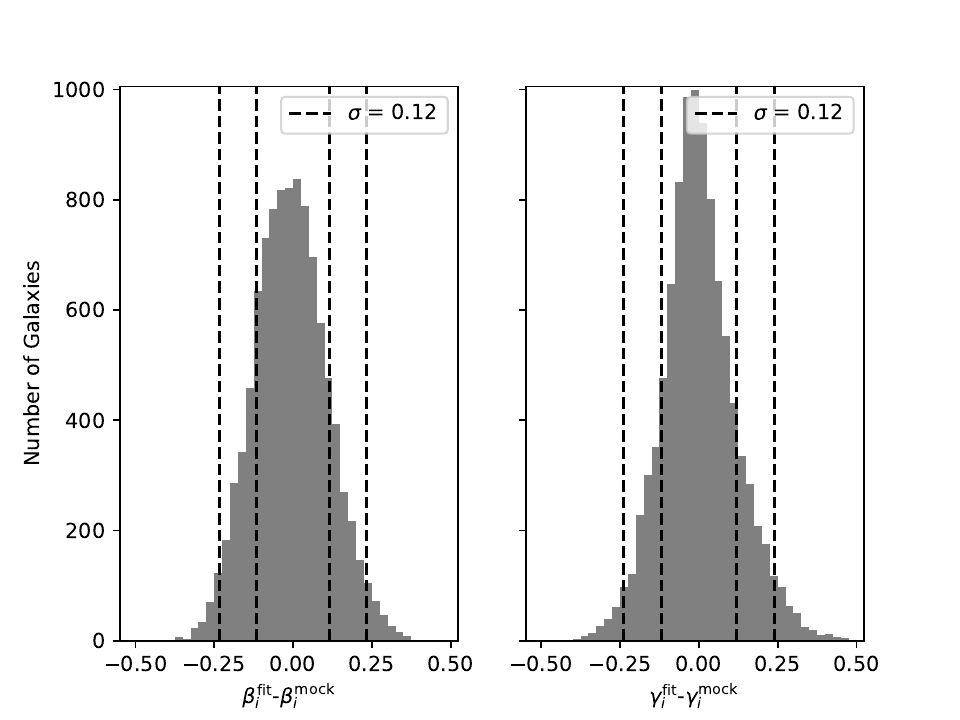}
    \caption{Histograms of the difference between input and recovered individual nuisance parameters. $\gamma^\mathrm{fit}_i$ and $\beta^\mathrm{fit}_i$ are the means of the posteriors for each individual lens. The standard deviation of both fitting errors are comparable than the intrinsic scatters of the population.}
    \label{fig:indi_error}
\end{figure}

Figure \ref{fig:cdmcorner} shows the posterior distributions of the $\Lambda$CDM, $w$CDM, and o$w$CDM models. The marginalized results are also shown in the figure. The parameters of lens populations are accurately recovered within 68$\%$ confidence level. Especially, the mean of $\gamma$ is constrained with high precision ($\pm$ 0.03 level). Figure \ref{fig:indi_error} shows the offset between the recovered and input values of the nuisance parameters in our model, i.e. the $\gamma_i$ and $\beta_i$ The standard deviation of the differences are both 0.12. This number is smaller than the intrinsic scatter of $\gamma$ and $\beta$ (0.16 and 0.13), and demonstrates that the model is accurately recovering these nuisance parameters as well as the population and cosmological parameters that we are primarily interested in. 

Every model successfully recovers the input cosmological parameters. For $w$CDM, we find $w = -0.98 \pm 0.11$. This is more precise than current individual cosmological probes: the Pantheon Type Ia supernovae data find $w = -0.90 \pm 0.14$ \citep{Brout2022}; baryon acoustic oscillations (BAO) from eBOSS give $w = -0.69 \pm 0.15$ \citep{eboss2021}. However, our results are less constraining on $w$ than the combination of both datasets and cosmic microwave background data from Planck, which yields $w$ = -1.03 $\pm$ 0.03 \citep{2020A&A...641A...6P}.

We find some strong covariances between the cosmological parameters and the lens population parameters of our hierarchical model. The most noticeable are covariances between $\langle\beta\rangle$ with $\omega_M$, $\Omega_k$ and $w$. This degeneracy arises from the fact that when mass is held constant, varying $\langle\beta\rangle$ leads to different velocity dispersions and, consequently, requires different cosmological parameters. Similarly, the mean density profile slope, $\langle\gamma\rangle$, is covariant with cosmology, since it governs the mapping from velocity dispersion to Einstein radius.

\begin{figure}
    \centering
    \includegraphics[width=0.5\textwidth]{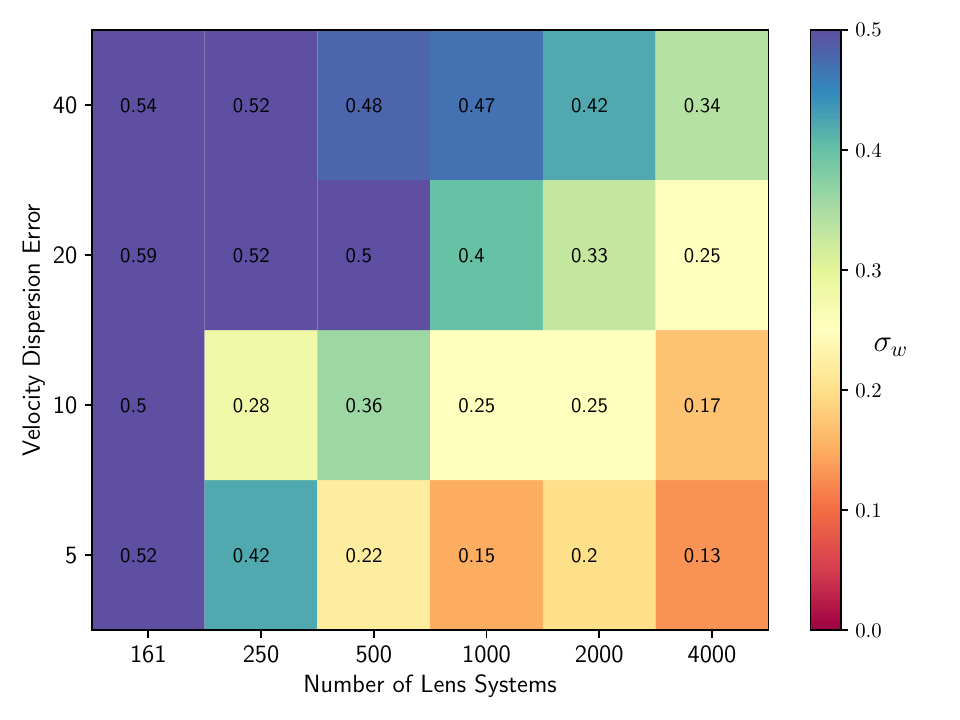}\\
    \includegraphics[width=0.5\textwidth]{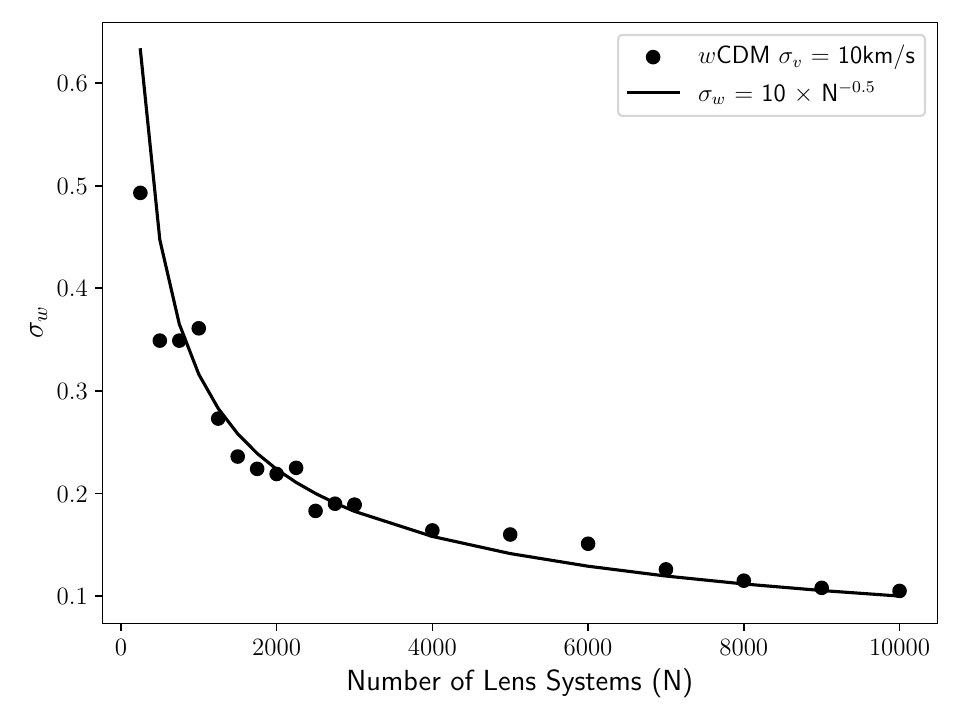}
   
    \caption{Top: 1 $\sigma$ uncertainty on the equation of state of dark energy as a function of numbers of lenses and observational velocity dispersion measurement uncertainty.\\
    Bottom: At a fixed LOSVD error of 10 km/s, the standard deviation of $w$ as function of number of lenses. The black dots represents the model result compared against the solid curve  of 10 times the square root of the number of lenses.}
     \label{fig:test_telescope}
\end{figure}

\subsection{The impact of data quality on the cosmological constraining power of lens samples}

The previous section was based on the assumption that 4MOST delivers 10 000 strong lenses with a velocity dispersion measured to a precision level of 10 km/s. In practice, it is difficult to accurately constrain the velocity dispersion of the lens galaxy in a spectroscopic survey. First, not all lens galaxies will be bright enough for their spectra to have sufficient signal-to-noise ratio (SNR) to measure with 4MOST \citep{2023Msngr.190...22I}. Second, measuring velocity dispersions from spectra is intrinsically hard, since we do not perfectly know the correct choice of stellar templates for the spectral energy distributions of lens galaxies \citep{collett2018,2017ApJ...845..157N}. Thirdly, continuum contamination from the source galaxy may be challenging to remove (but see Turner et al 2023, submitted)

In this section, we explore how the constraining power for the $w$CDM cosmological model varies when we have different sample sizes or velocity dispersion measurement errors. The results are shown in the top panel of figure \ref{fig:test_telescope}. As expected, the constraining power on $w$ improves as we increase the number of samples or improve the velocity dispersion measurement. The $w$ constraint versus sample size roughly follows $\sigma_w = 10 \times  N^{-0.5}$ as shown in the bottom panel of figure \ref{fig:test_telescope}. For 4MOST observations, there will be a trade-off between observing more lens systems with relatively low SNR (resulting in larger LOSVD errors) and obtaining longer exposure time for each target (resulting in higher SNR but lower sample size). We find that generally, decreasing the LOSVD error by a factor of two (improving $w$ by a factor of $\approx$ 1.3-1.7) is comparably helpful for $w$ as increasing the sample size by a factor of two.

As shown in figure \ref{fig:indi_error}, the individual $\beta_i$ are constrained comparably well to direct measurements of the circular velocity curve using HST \citep[0.12 vs 0.05-0.4,][]{2001AJ....121.1936G}. The individual $\gamma_i$ error (0.12) is about 5 times larger than the typical result from lens modeling (e.g., 2.11 $\pm$ 0.04 by \cite{Dyewarren2005}, 1.96 $\pm$ 0.02 by \cite{Dye2008}, and 2.08 $\pm$ 0.03 by \cite{Suyu2010}). This indicates that adding constraints on individual $\gamma_i$ through lens modeling, instead of leaving them as nuisance parameters would substantially improve the cosmological constraining power.

\begin{figure}
    \centering
    \includegraphics[width=0.45\textwidth]{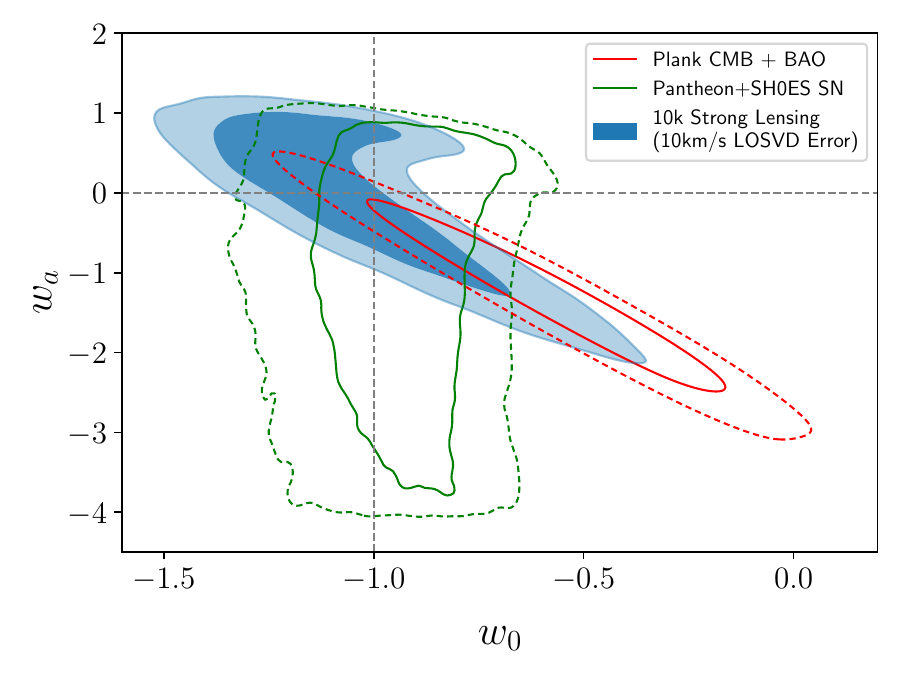}
    \caption{Constraints on a flat cosmological model with evolving $w$ where $w$ = w$_0$ + w$_a$(1-a). The grey dashed line is the fiducial value of our mocks. The blue contour is the posteriors distribution of $w_a, w_0$ with 10 000 strong lensing systems. The green and red lines show current results from Type Ia supernova and CMB+BAO, respectively \citep{2022ApJ...938..110B, 2020A&A...641A...6P}.}
    \label{fig:my_waw0}
\end{figure}

\subsection{Evolving Equation of State of Dark Energy}

The equation of state of dark energy ($w$) is defined as the ratio of pressure over energy density. In many theoretical cosmological models, $w$ evolves with redshift (e.g. \cite{1988ApJ...325L..17P, 2002PhLB..545...23C, 2005PhLB..607...35F}). We test our method on an evolving $w$ where $w(z)$ = $w_0$ + $w_a \frac{z}{1+z}$ \citep{1998Ap&SS.261..303C}. Since our mock data has fixed $w = -1$, the fiducial value this model is $w_0 = -1$, $w_a = 0$ \citep{2001IJMPD..10..213C, PhysRevLett.90.091301}. 

Our forecast constraints for 10 000 lenses are shown in Figure \ref{fig:my_waw0}. The 1D marginalized posteriors give $w_a = 0.0^{+0.8}_{-1.4}$, $w_0 = -1.0^{+0.5}_{-0.4}$. Since these parameters are strongly covariant, we use the figure of merit (FoM) from equation 6 in \cite{PhysRevD.82.063004} to quantify the overall dark energy constraining power:

\begin{eqnarray}
\mathrm{FoM}=\frac{6.17 \pi}{A_{95}},
\end{eqnarray}

Where $A_{95}$ is the Area enclosed within the 95$\%$ confidence contour in the $w_0-w_a$ plane. Larger FoM's therefore imply better dark energy constraining power. The FoM of our 10 000 lens result is 15. As a comparison, the 10-year LSST forecast for the combination of weak lensing and Large Scale Structure which has a FoM of 49 \citep{2018arXiv180901669T}, for LSST 10-year supernovae the FoM is expected to be 32.  We find that the galaxy-galaxy strong lensing constraints have very similar $w_0-w_a$ covariance as those forecast for weak lensing. This is to be expected since both methods constrain the distance ratio $D_{s} / D_{ls}$.

\section{The impact of the evolution of Lens Galaxy Population}
\label{sec:evolving}
The results of the previous section were based on mock data where $\gamma$ and $\beta$ follow Gaussian distributions that do not evolve with redshift. However, this assumption may not be true of the real Universe. Lens galaxies might have evolving density slopes due to various reasons. At high redshift, dissipative processes like gas-rich mergers dominate, leading to steeper total density slopes (large $\gamma$, see \cite{2013ApJ...766...71R, 2014ApJ...786...89S}). At redshifts lower than $z \sim 2$ (which is all the lenses in our sample), dissipationless processes like gas-poor mergers can flatten the density slopes (e.g., \cite{2012MNRAS.425.3119H, 2013MNRAS.429.2924H}). 

Hydrodynamic simulations show that the mass-density slope of early-type galaxies (ETGs) becomes flatter at lower redshift \citep{2012ApJ...754..115J, 2013ApJ...766...71R, 2017MNRAS.464.3742R, 2020MNRAS.491.5188W}. However, most strong lensing studies find no evidence for evolving density slopes with redshift \citep{Koopmans:2005ig, 2010ApJ...724..511A, 2011ApJ...727...96R, 2012ApJ...757...82B, 2011MNRAS.415.2215B, 2017SCPMA..60h0411C}, or a slight decrease in density slopes with redshift \citep{2013ApJ...777...98S, 2015ApJ...806..185C, 2016MNRAS.461.2192C, 2017MNRAS.471.3079H}. This might suggest that strong lensing surveys are biased towards certain lens populations, or that there is no evolution at all at low redshift. Since the possibility of an evolving mass profile has not been ruled out, we explore the robustness of our method when applied to an evolving lens population.

According to the simulations of \cite{2017MNRAS.464.3742R}, the evolving total mass profile can be approximated by a linear relation: $\gamma_{z} = 0.21z + 2.03$. This equation gives unreasonably steep profiles compared to observed lenses, so for our new mock data we use the evolution of \cite{2017MNRAS.464.3742R}, but fix the mean and scatter to give a total population that has the same average slope and scatter as in Section \ref{sec:results}. Thus we draw our density slopes from $\langle \gamma_{z} \rangle = 0.21z + 1.89$ with a scatter of 0.155.

\begin{figure}
    \centering
    \includegraphics[width=0.5\textwidth]{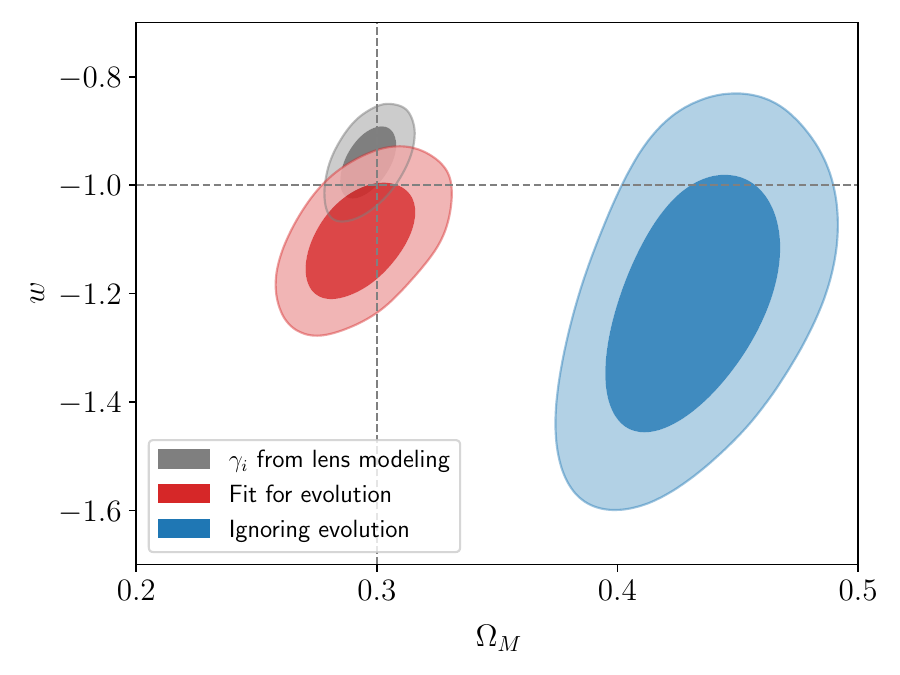}
    \caption{The impact on the inferred $w$ and $\Omega_\mathrm{M}$ of redshift evolution in the mean lens density profile. The blue contours represent a model that does not allow for redshift evolution in $\gamma$. The red contours represent the result for a model which fits for linear evolution of the mean population $\gamma$. The grey contour results when individual $\gamma_i$ are measured through detailed lens modelling of every lens.}
    \label{fig:pre_measure}
\end{figure}

We first try to fit the redshift evolving lens dataset with the non-evolving hierarchical model used in  Section \ref{sec:results}. The resulting cosmology posteriors distributions for 10 000 lens systems are shown in blue in figure \ref{fig:pre_measure} (full result in appendix \ref{appendix}). The population parameter of $\langle\gamma\rangle$ population is recovered, but the cosmology parameters and the stellar anisotropies are systematically incorrect. Thus, a more complex model is needed to deal with this scenario.

We can generalize our hierarchical model to account for population redshift evolution by fitting additional parameters that describe the evolution of the population. In order to keep our parameters as similar as possible to those in section \ref{sec:results}, we parameterise our density profile redshift evolution as follows:
\begin{equation}
   \gamma_z =  \langle\gamma\rangle + \Delta_\gamma \times (z - 0.47) \pm \sigma_\gamma. 
\end{equation}
where 0.47 is the mean lens redshift of our population, we set $\langle \gamma \rangle = 2$ as the ensemble mean density profile slope and $\Delta_\gamma = 0.21$ as its linear evolution with redshift.

Unlike the model where we ignore the evolution of the population, we find that fitting for the evolution does reproduce the input lens galaxy population parameters (Figure \ref{fig:appendix1}). We recover the correct redshift density evolution: $\Delta_\gamma= 0.195^{+0.015}_{-0.017}$. Using this method, we are able to constrain $w$ to -1.1 $\pm$ 0.07 level (Figure \ref{fig:pre_measure}), which is better than the $\pm$ 0.11 error that we found for $w$CDM without $\gamma$ evolution. For a flat $w_0$$w_a$CDM cosmology, the figure of merit improves to 17 as shown in the top panel of the figure \ref{fig:gammawaw0cdm}. The modest improvements are due to the slightly reduced intrinsic scatter of our mock population.

Alternatively, the population parameters \st{of} describing density profiles can be constrained on a lens-by-lens basis using detailed lens modelling of the arcs observed in each system, 
without the need to know the cosmological model. This is because the slope of the mass density determines the radial derivative of the deflection angles, and thus the radial width of the arc is sensitive to the density profile slope \citep{Dyewarren2005, Suyu2010, collett2018}. With image quality comparable to that of HST and Euclid, one can constrain $\gamma$ to a precision level of $\sigma_\gamma \approx 0.02$ \citep{2015JCAP...09..059M}, although this neglects the impact of the mass-sheet degeneracy \citep{Gorenstein1988,2000AJ....120.1654S, 2002MNRAS.332..951W, 2012MNRAS.425.1772L, Schneider&Sluse2013}. Whilst lens modelling at scale is currently challenging, adding a precise prior on $\gamma_i$ for each lens should significantly improve the constraints on cosmological parameters. If we assume that we are able to pre-determine the value of $\gamma$ for each lens system with a precision level of 0.02, we can treat $\gamma$ as an observable, similar to the Einstein radius. The measurement of "$w$" is greatly improved in this scenario: $w = -1.01 \pm 0.06$ for a 10 000 lenses with no evolution of the population, or $w = -1.08 \pm 0.07$ for the population where $\Delta_\gamma= 0.21$. Respectively, the figure of merit improves to 28 and 64 for 10 000 lenses assuming a flat $w_0$$w_a$CDM cosmology (see the bottom two panels of the figure \ref{fig:gammawaw0cdm}). 

\section{Larger scatter on lens population}

In this study, our mock lenses are created, assuming the distribution of $\sigma_\gamma$ and $\sigma_\beta$ match those inferred from SLACS data \citep{2010ApJ...724..511A} and a set of nearby elliptical galaxies \citep{2001AJ....121.1936G}. This assumption is critical to our forecasts since $\sigma_\gamma$ and $\sigma_\beta$ quantify how standardizable each lens is. If $\sigma_\gamma$ or $\sigma_\beta$ are significantly larger in the real Universe then the power of this method to constrain cosmography will be greatly diminished. It should be noted that the lens systems were selected from ground-based sky surveys, which may introduce biases compared to space-based surveys like Euclid. Additionally, the nearby elliptical galaxies might not be a perfect representation of lens elliptical galaxies. For instance, \cite{2017MNRAS.469.1824X} analyzed elliptical galaxies in the Illustris simulation and found that the standard deviation of velocity anisotropy can reach 0.3, while in our work, we employed a value of 0.13. 

We generate mock galaxies with a wider range of $\gamma$ (0.1 - 0.3) and $\beta$ (0.1 - 0.3) to test the effectiveness of our method. Table \ref{tab:large_standard deviation} presents the constraints on $w$ ($\sigma_w$) with 1,000 lens systems in a flat universe under different distributions of $\gamma$ and $\beta$. The constraints deteriorate as the scatter in either $\gamma$ or $\beta$ increases, but even in the most pessimistic scenario the constraints degrade by a factor of 2.17. On the other hand, our forecasts may be pessimistic if lenses can be further standardized by understanding the physical properties that drive the scatter of $\gamma_i$ and $\beta_i$ away from the population mean.

\begin{table}
    \centering
    \begin{tabular}{c|ccc}
        \hline\\
         &$\mathrm{\sigma}_{\gamma}=0.1$ & $\sigma_{\gamma}=0.2$ & $\sigma_{\gamma}=0.3$ \\
        \hline\\
        $\sigma_{\beta}=0.1$& 0.90 & 1.03 & 1.33 \\
        $\sigma_{\beta}=0.2$& 1.15 & 1.26 & 1.47  \\
        $\sigma_{\beta}=0.3$& 1.52 & 1.79 & 2.17  \\
        \hline\\
    \end{tabular}
    \caption{The relative change of the 68 percent confidence interval of $w$ as a function of how standardizable lenses are. We assume with 10 000 lens systems, but change the intrinsic scatter of the lens population $\gamma$ and $\beta$.}
    \label{tab:large_standard deviation}
\end{table}

\section{Application on existing data}

\label{sec:161lens}
\begin{figure*}
    \centering
    \includegraphics[width=0.9\textwidth]{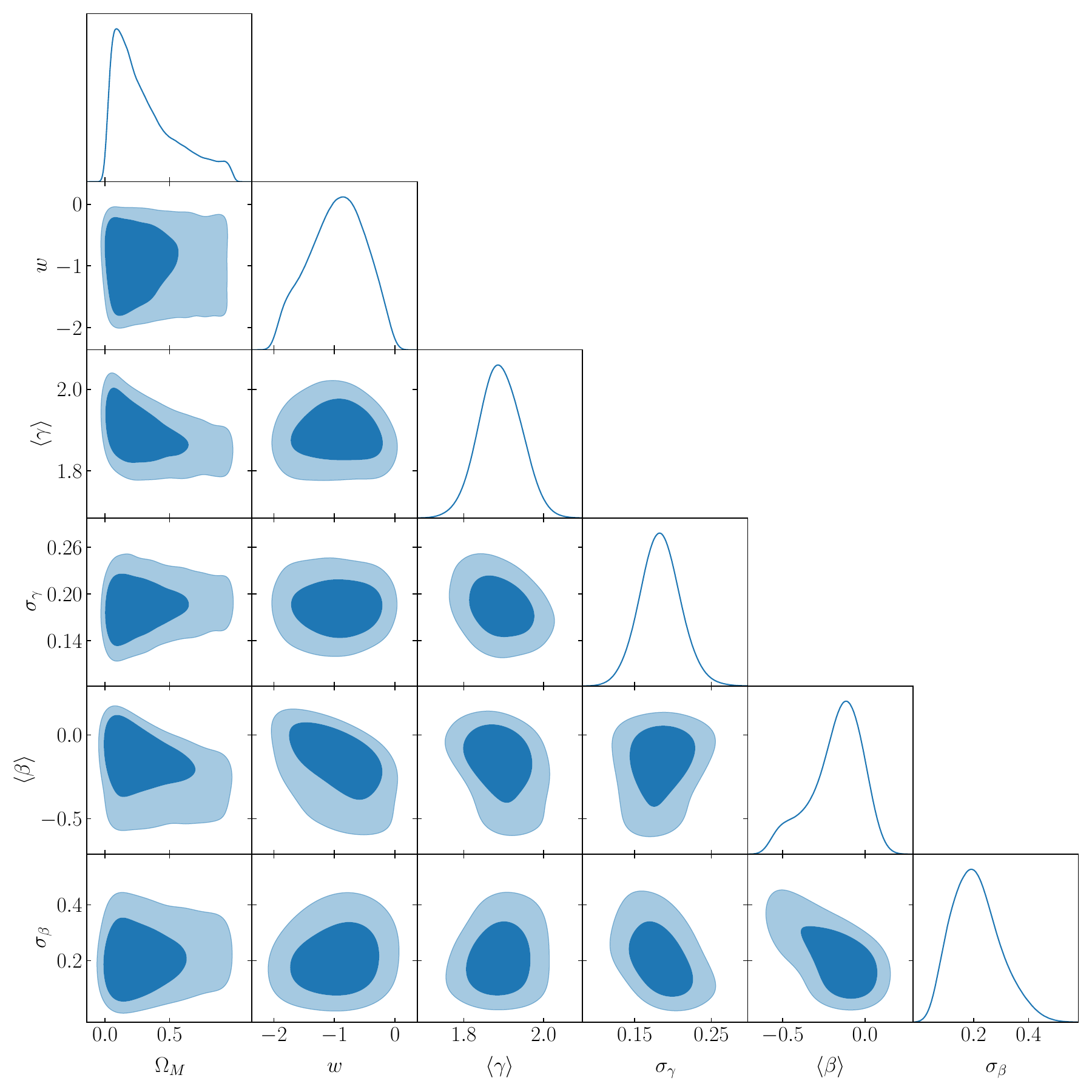}
    \begin{tabular}{c|c|c|c|c|c|c|c|c}
        \hline
        Parameter & $\Omega_M$ & $w$ & $\gamma$ & $\sigma_{\gamma}$ & $\beta$ & $\sigma_{\beta}$ \\
        \hline
        $w$CDM & 0.32 $\pm$ 0.25 & -0.96 $\pm$ 0.46 & 1.89 $\pm$ 0.05 & 0.18 $\pm$ 0.03 & -0.17 $\pm$ 0.16 & 0.21 $\pm$ 0.09\\
        \hline
    \end{tabular}
    \caption{Posterior distribution and 1D marginalized distribution of the cosmological and lens population parameters assuming a $w$CDM universe and 161 real lenses from \citet{Cao_2015} and \citet{Chen:2018jcf}.}
    \label{fig:salcs161}
\end{figure*}

As a simple example of our method, we apply it to existing data to evaluate how well our power-law mass model can describe real lens galaxies. Additionally, we can compare our results with previous research that used a similar method and the $w$CDM model \citep{2015ApJ...806..185C, Chen:2018jcf}. We use a dataset of 161 lens systems selected by \cite{Chen:2018jcf} (see Table A1), which were obtained from surveys including LSD, SL2S, SLACS, and S4TM. They measured the luminosity density slope and calculated the equivalent fiber radius of the spectrum for each galaxy. As we lack the specific $\delta$ value for each individual galaxy, we treat $\delta$ as an unknown value and draw from its measured distribution: $\delta = \mathcal{N}(2.173, 0.085)$. The average velocity dispersion error for this dataset is 22 km/s.

The resulting posterior distribution is shown in figure \ref{fig:salcs161}. In our results, we find that $w$ = -0.90 $\pm$ 0.45, which is an even better measurement compared to the theoretical result shown in Figure \ref{fig:test_telescope}. Regarding galaxy population properties, we obtain $\langle\gamma\rangle = 1.89\pm+0.05$, $\sigma{\gamma} = 0.18\pm0.03$, $\langle\beta\rangle = -0.17\pm0.16$, and $\sigma_{\beta} = 0.21\pm0.09$ at a 68$\%$ confidence level. The predicted $\gamma$ and $\beta$ populations are both smaller than those reported in other studies. This is likely due to the fact that we lack accurate $\delta$ values. Additionally, these 161 lens systems were obtained from four different surveys, and the population (selection function) of lens systems from different surveys can vary in which case fitting the ensemble with a single lens population would be incorrect.  Understanding the selection function between different surveys is crucial for accurately measuring the lens population because the future strong lensing sample is likely to be the combination of LSST and Euclid discoveries.

\section{Discussion and conclusion}

In this work, we have constructed a hierarchical model of galaxy-galaxy lenses and the underlying cosmological parameters. We have employed Hamiltonian Monte Carlo through JAX-based NumPyro modelling to efficiently perform the analysis. Our findings indicate that we can simultaneously constrain cosmological parameters and lens galaxy properties under different cosmological models, including $\Lambda$CDM, $w$CDM, and o$w$CDM. With a sample size of 10 000 lens systems, our method should achieve a 68$\%$ confidence interval of $\pm$0.11 on $w$. These levels of constraint are comparable to other cosmic probes such as the Cosmic Microwave Background (CMB), standard candles, weak lensing, and galaxy clustering. Furthermore, we have shown that the evolution of the lens population can also be simultaneously constrained. We also tested the ability of galaxy-galaxy lenses to constrain evolving dark energy. Our forecast Figure of Merit is 15, which is within a factor of 3 of each individual probe forecast for the LSST 10-year cosmological constraints \citep{PhysRevD.82.063004}. These results rely only on measurements of the Einstein radius, redshifts, velocity dispersion, and luminosity profile of each lens. Additional constraints on the lens density profile from detailed lens modelling, would improve the cosmological constraining power by a factor of $\sim$2, however, our method can still work without this huge investment in detailed modelling.

Additionally, we applied our model to 161 real lenses in Section \ref{sec:161lens}, finding $w=-0.90 \pm 0.45$. As we discuss below, this result is likely systematics dominated but it illustrates that the method does work on real data.

One of the major simplifications made in this study is that the mass and light models used may not accurately represent real galaxies. The power law light profile we employ is a simplification of the more general Sersic profile \citep{1963BAAA....6...41S}, which can describe the light profile of most elliptical galaxies. The effect of the PSF is also not considered in our model, which will bring extra complexity to the velocity dispersion measurement. Furthermore, galaxies are made of dark and luminous matter, and whilst the 'bulge-halo conspiracy' gives total density profiles that are close to powerlaw, the absolute truth is undoubtedly more complex than we have assumed. Our powerlaw assumptions, make the mathematics of our problem analytic (Equation \ref{eqa:main}), but should not hugely change the final constraints, since the model is only relevant in mapping the aperture dynamical mass onto the mass within the Einstein radius.

In this study, we made the assumption that all galaxy properties follow Gaussian distributions, this might not be true in reality. Additionally, due to a lack of observational evidence, we assumed that there are no correlations between $\gamma$, $\beta$, and $\delta$. However, in actuality, these properties are likely to have initial scaling relations. For example, \cite{2010ApJ...724..511A} found that a steeper mass density profile implies a higher central surface mass density. Furthermore, \cite{2007MNRAS.379..418C} discovered that elliptical galaxies can be categorized as slow and fast rotators, and they exhibit different $\beta$ populations. There are also some outliers that have very low $\beta$ found by \citep{2001AJ....121.1936G}.  In our results, the orbital velocity anisotropy of lens galaxies exhibits a strong degeneracy with cosmological parameters. Therefore, understanding these correlations will aid in better fitting the mass models of individual galaxies. Another potential observational bias arises from our assumption that elliptical galaxies have a constant $\beta$, whereas observations suggest that elliptical galaxies have varying $\beta$ with radius. Specifically, the central region of typical massive elliptical galaxies tends to be isotropic or mildly radially anisotropic \citep{2001AJ....121.1936G, 2007MNRAS.379..418C}. In spectroscopic surveys, the fiber has a fixed radius, which means that the region from which we measure the velocity dispersion will have a different size depending on the half-light radius. Consequently, any observed evolution in the properties may simply be a side effect resulting from the evolution of the lens galaxy's angular size. Similarly, we have ignored the potential for redshift evolution in the population anisotropy, which may be expected from simulations \citep{2017MNRAS.469.1824X}.

The simulated lens population in this study was limited to a redshift cutoff of 1.5. In real observations, the lens population is subject to additional limitations. For instance, observing the lens galaxy and the source galaxy simultaneously in a spectroscopic survey might not be feasible for lens systems with a large Einstein radius. As a result, some low-redshift lens galaxies may be ruled out from the analysis. Additionally, faint lens galaxies can lead to large velocity dispersion measurement errors, making them less likely to be included in the analysis. This can exclude low-mass galaxies and consequently affect the $\gamma$ population estimation. Conversely, luminous galaxies beyond the redshift cutoff might be included in the analysis, which can introduce further complexities. These observational limitations and selection effects should be taken into account when interpreting the results and considering the true lens galaxy population.

The method used in this study can also be valuable for the analysis of time-delay cosmography, such as gravitational lens quasars and supernovae. Similar to our approach, galaxy density profiles can be simultaneously inferred with the Hubble constant, as demonstrated in previous work (e.g., \cite{2020A&A...643A.165B}). But their mass models are heavily prior dominated, the bias in $\gamma$ will introduce a significant bias in $\mathrm{H}_0$ in LSST era where the number of lensed quasar/SNe are large \citep{2016MNRAS.462.3255C}. The lens galaxy population parameters we obtained can serve as priors when modeling gravitational lens quasars or supernovae. However, it is important to note that normal lens systems may have different population parameters compared to lensed quasar or supernova systems. Conducting a detailed investigation into the selection function between these systems will be necessary to mitigate any biases.

Whilst this work is a simplified investigation of how to simultaneously constrain the astrophysics of lenses and the underlying cosmological parameters, it has shown that the potential constraining power is competitive to more established cosmological probes. These encouraging results motivate further work on modelling the selection function, fitting more general density and dynamical profiles, and gathering the required data.

\section*{Acknowledgements}
We are grateful to Andy Lundgren, Giovanni Granata, Hannah Turner, Simon Birrer, Sydney Erickson, Shuo Cao, Shude Mao, Shawn Knabel, Phil Marshall, Russell Smith, and Leon Koopmans for helpful conversations that have enriched this work.

Numerical computations were done on the Sciama High Performance Compute (HPC) cluster which is supported by the ICG, SEPNet, and the University of Portsmouth.

This work has received funding from the European Research Council (ERC) under the European Union's Horizon 2020 research and innovation programme (LensEra: grant agreement No 945536). TC is funded by the Royal Society through a University Research Fellowship.
For the purpose of open access, the authors have applied a Creative Commons Attribution (CC BY) license to any Author Accepted Manuscript version arising.

\section{Data Availability}
Forecast lens population data are available at \href{github.com/tcollett/LensPop}{https://github.com/tcollett/LensPop}. Model posterior chains are available from the corresponding author on request. The parameters of 161 Lens systems are available at \cite{Chen:2018jcf} in \href{https://doi.org/10.1093/mnras/stz1902}{https://doi.org/10.1093/mnras/stz1902}.



\bibliographystyle{mnras}
\bibliography{CosmologywithLensing} 

\appendix
\section{}
\label{appendix}
The appendix contains 2D distributions that are not shown in the main body.

\begin{figure}
    \centering
    \includegraphics[width=0.45\textwidth]{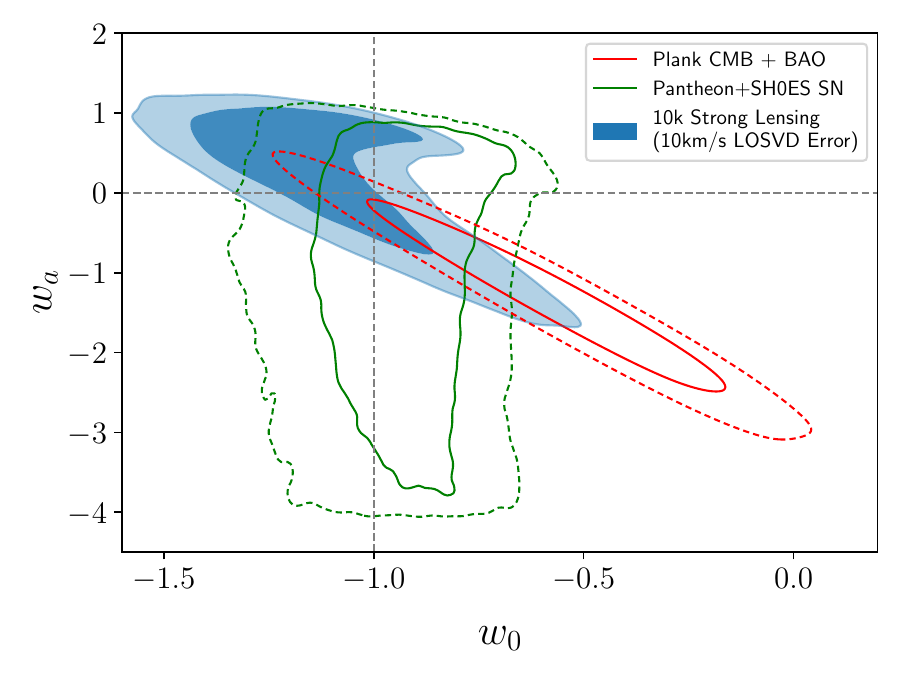}
    \includegraphics[width=0.45\textwidth]{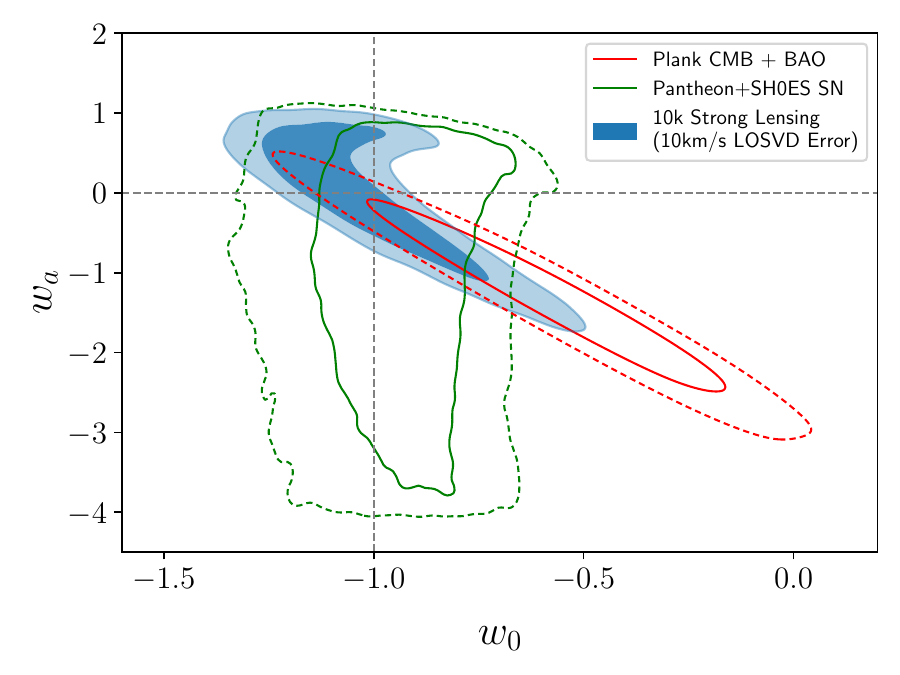}
    \caption{Constraints on $w_0$ and $w_a$ for a redshift evolving population mean density profile slope. The top panel results for a model which fits for linear evolution of the mean population $\gamma$ (FoM=17). The bottom panel results when individual $\gamma_i$ are measured through detailed lens modelling of every lens (FoM=28).}
    \label{fig:gammawaw0cdm}
\end{figure}

\begin{figure*}
    \centering
    \includegraphics[width=0.9\textwidth]{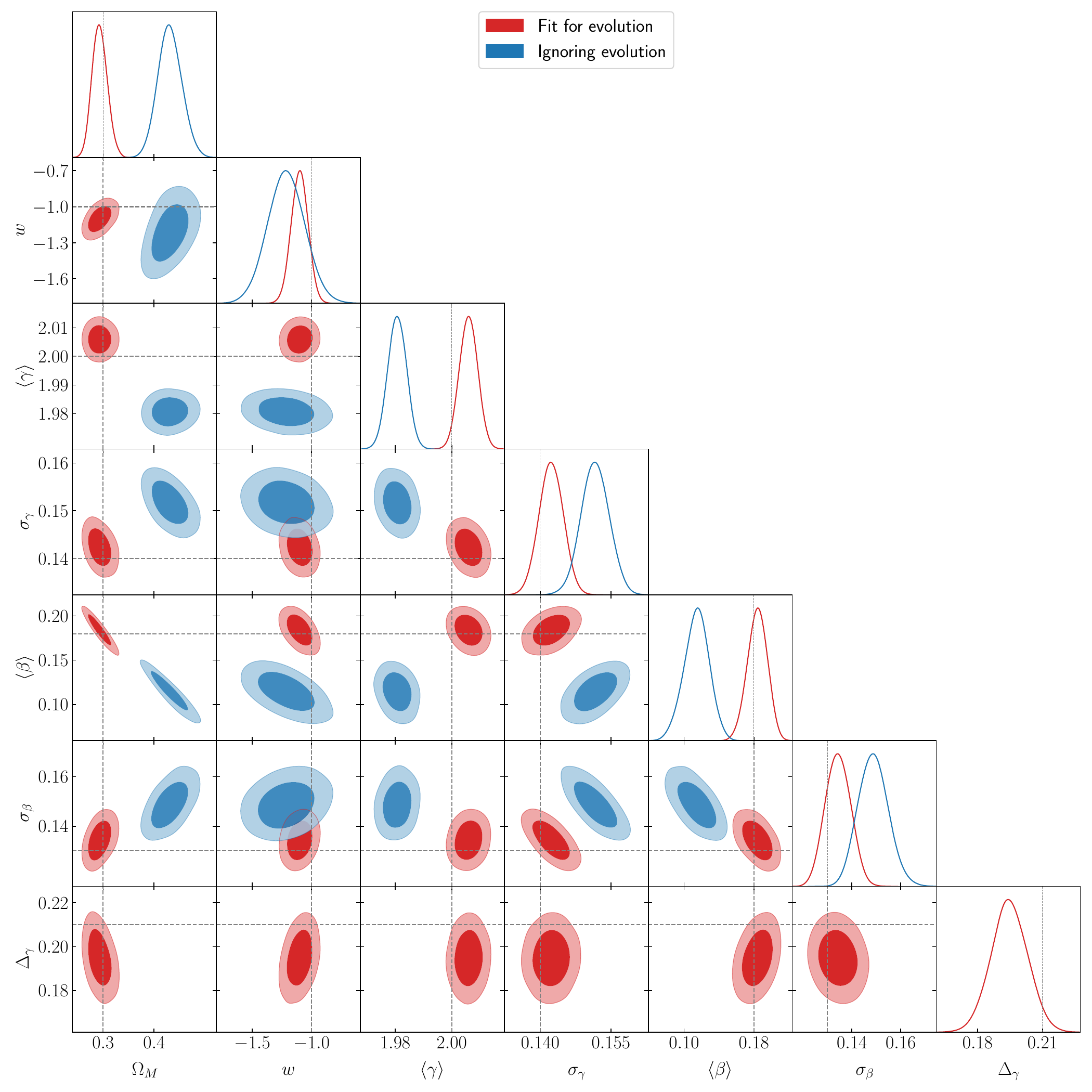}
    \caption{The full 2D posterior distributions obtained for a redshift evolving population mean density profile slope. The blue contours result from neglecting redshift dependence of $\gamma$ whilst the red contours allow $\gamma$ to evolve linearly with redshift. }
    \label{fig:appendix1}
\end{figure*}



\bsp	
\label{lastpage}
\end{document}